\documentclass[aps,preprint,showpacs,showkeys]{revtex4}
\usepackage[pdftex]{graphicx}
\usepackage{amssymb}
\usepackage{bm}

\begin{document}
\setcounter{page}{1}
\title[]{Propagation of Light in Doubly Special Relativity}
\author{Sung Ku \surname{Kim}}
\affiliation{Department of Physics, Ewha Woman's University, Seoul
  120-750, Korea}
\author{Sun Myong \surname{Kim}}
\email{smkim@dragon.yonsei.ac.kr}
\affiliation{Department of Physics, Yonsei University, Wonju, Kangwon-Do
220-710, Korea}
\author{Chaiho \surname{Rim}}
\affiliation{Department of Physics, Chonbuk National University,
  Chonju 561-756, Korea}
\author{Jae Hyung \surname{Yee}}
\affiliation{Institute of Physics and Applied Physics, Yonsei
  University, Seoul 120-749, Korea}
\date[]{Received January 5 2004}

\begin{abstract}
In an attempt to clarify what is the velocity of a particle in
doubly special relativity, we solve Maxwell's equations invariant
under the position-space nonlinear Lorentz transformation proposed
by Kimberly, Magueijo, and Medeiros. We show that only the
amplitude of the Maxwell wave, not the phase, is affected by the
nonlinearity of the transformation. Thus, although the Maxwell
wave appears to have infinitely large energy near the Planck time,
the wave velocity is the same as the conventional light velocity.
Surprisingly, the velocity of the Maxwell wave is not the same as
the maximum signal velocity determined by the null geodesic
condition, which is infinitely large near the Planck time and
monotonically decreases in time to the conventional light velocity
when time approaches infinity. This implies that, depending on the
position of the particle in question, the light cone determined by
Maxwell's equations may be inside or outside the null cone
determined by the null geodesic equation, which may lead to the
causality problem.
\end{abstract}



\maketitle

\section{Introduction}

Doubly special relativity (DSR), which is a generalization of
special relativity with two observer-independent scales, has
attracted much attention recently in an attempt to describe the
phenomenological aspects of physics on the Planck scale. It was
first introduced as a generalization of the Lorentz
transformation, respecting the principle of relativity in a linear
fractional form [1-3]. It was recently suggested again as a
generalization of special relativity to include one more invariant
scale, in addition to the speed of light, introduced in a modified
Lorentz transformation in momentum space\cite{dsr-intro},  and it
was subsequently shown to be a nonlinear representation of the
Lorentz group [5,6].
It has also been pointed out that these theories can be understood
as particular bases of the $\kappa$-Poincare theory based on
quantum (Hopf) algebra, and the resulting energy-momentum space
admits description in terms of a deSitter-type
geometry\cite{novak}. Although DSR defined in momentum space has
provided a possibility of explaining phenomena such as
ultra-high-energy cosmic rays with energies above the
Greisen-Zatsepin-Kuzmin (GZK) cut-off and high-energy gamma rays
[4,5,8]
and a candidate for
a varying light velocity theory for cosmology\cite{gray}, some
questions on the interpretation and the consistency of the
proposal have been raised.

First of all, it is not clear how to define the velocity of a
particle unambiguously in an inertial frame\cite{iframe}. The
authors of Ref. 10 has also raised the possibility of large-scale
non-locality in the DSR defined in momentum space. The root of
such problems lies in the fact that we do not know the exact
position-space transformation corresponding to the modified
Lorentz transformation defined in momentum space, such as those
suggested in Refs. 4 and 5.
In position space, the velocity of a particle is unambiguously
defined as $dx^i/dt$ and the unique momentum of the particle from
the velocity. Also, the (non)locality of field theory can be
addressed unambiguously in position space whereas in Ref. 10,
a Fourier transformation was used to obtain the behavior of the
field in position space. To use a Fourier transformation, one has
to rely on the quantum-mechanical relation
$p_i=-i\partial/\partial x^i$ for discussion of classical theory.
Although there have been attempts to obtain the position-space
transformation corresponding to the given momentum-space nonlinear
Lorentz transformations [11,12],
the results depend on the momentum, and their physical
interpretation is not yet clear.

Thus, if the physical implications of nonlinear relativity are to
be understood, it is preferable to start with the modified Lorentz
transformation defined in position space. The nonlinear
representation of the Lorentz group in position space with a large
invariant length scale was first proposed by Fock\cite{fock}, the
so-called Fock-Lorentz transformation, and elaborated by
Manida\cite{manida} and Stepanov\cite{stepanov}. Recently,
Kimberly, Magueijo, and Medeiros (KMM) proposed a nonlinear
relativity in position space with a Planck scale
invariant\cite{kmm}, which is relevant in applications to
Planck-scale physics. The purpose of this paper is to study the
physical implications of nonlinear relativity theories defined in
position space.

In the next section, we study the space-time structure of the
position-space nonlinear relativity proposed by KMM. We show that
the maximum signal velocity determined by the null geodesic
condition is infinitely large near the Planck time and
monotonically decreases in time to the light velocity, c, when
time becomes infinite. In Section III, we solve Maxwell's
equations invariant under the KMM transformation proposed in Ref.
11. Although the Maxwell wave appears to have infinitely large
energy near the Planck time, its velocity is shown to be the same
as the conventional light velocity. Finally, in Section IV, we
discuss some implications of our results.

\section{Space-time Structure of Nonlinear Relativity in Position
Space}

In this section, we study the space-time structure of the
nonlinear relativity proposed by KMM\cite{kmm} and will briefly
discuss that of the Fock-Lorentz case [1-3].
The nonlinear realization of the Lorentz group may be obtained by
means of a transformation $V$ such that the new boost generators of
the Lorentz group with respect to space-time coordinates are
\begin{eqnarray}
K^i=V^{-1} L_{0}^{~\:i}~ V, \label{boost}
\end{eqnarray}
where
\begin{eqnarray}
L_{\alpha\beta}=x_{\alpha}\frac{\partial}{\partial
x^{\beta}}-x_{\beta}\frac{\partial}{\partial x^{\alpha}}
\end{eqnarray}
are the standard Lorentz generators. Despite this transformation,
the new generators satisfy the ordinary Lorentz algebra,
\begin{eqnarray}
[J^i,K^j]=\epsilon^{ijk}K_k ,~~ [K^i,K^j]=\epsilon^{ijk}J_k ,~~
[J^i,J^j]=\epsilon^{ijk}J_k,
\end{eqnarray}
where $J^i=\epsilon^{ijk}L_{jk}$. Thus, the new generators
constitute a new (nonlinear) representation of the Lorentz
algebra. KMM chose $V$ as
\begin{eqnarray}
V=e^{\frac{t_{P}}{t}D}, \label{genl}
\end{eqnarray}
where $D=x^{a}\partial_{a}$ and $t_{P}$ represents a new invariant
scale, the Planck time.   Now, due to Eq. (\ref{boost}), the boost
generators for nonlinear relativity take the form
\begin{eqnarray}
K^{i}=L_{0}^{~\:i}-\frac{t_{P}x^{i}}{t^{2}}D,
\end{eqnarray}
which in turn induces a nonlinear representation of a Lorentz
transformation. The nonlinear Lorentz transformation is given by~
$x'^\mu=V^{-1} e^{\omega^{\alpha\beta}L_{\alpha\beta}} Vx^{\mu}$:
\begin{eqnarray}
t'&=&\gamma(t-\textit{v} x^{3})(1-\frac{t_{P}}{t})+t_{P},
\label{transt}\\
x'^3&=&\gamma(x^{3}
    -\textit{v}t)[(1-\frac{t_{P}}{t})
    +\frac{t_{P}}{\gamma(t-\textit{v}x^{3})}],
\label{transx3}\\
x'^a&=&x^{a}[(1-\frac{t_{P}}{t})+\frac{t_{P}}{\gamma(t-\textit{v}x^{3})}],~~a=1,2,
\label{transx12}
\end{eqnarray}
where  $\textit{v}=\tanh \omega^{03}$ is the relative velocity
between two frames, which is assumed to be in the $x^3$-direction,
and $\gamma=\frac{1}{\sqrt{1-\textit{v}^{2}}}$ is the Lorentz
factor. Note that the transformations in Eqs.
(\ref{transt})-(\ref{transx12})
are singular at $t=t_{P}$, the Planck time. Thus, the theory
describes the region of space-time after or before the Planck
time. We also note that changing the sign of $t_P$ (or $t$) in Eq.
(4) leads to another nonlinear representation of the Lorentz
group, so one can implement the time reversal symmetry in the
system, as noted by KMM \cite{kmm}. As KMM pointed out \cite{kmm},
the nonlinearity of the representation destroys translational
invariance. Thus, the addition law for the coordinate variables is
not obvious and needs further study, as is the case for the
momentum variables in DSR defined in momentum space. Even though
the position space DSR proposed by KMM and by FR nonlinear
relativity poses many problems, the position space DSR can be
useful in understanding how to define the velocity of light. To
this end, we will consider the null geodesic condition in this
section and the Maxwell equation in the next section.

The nonlinear KMM transformation, Eqs.
(\ref{transt})-(\ref{transx12}),
can be represented as a linear Lorentz transformation in terms of
the new variables
\begin{eqnarray}
\tilde{x}^\mu=x^\mu (1-\frac{t_P}{t}).
\end{eqnarray}
It is easy to check that the quantity
\begin{eqnarray}
\tilde{t}^2-\vec{\tilde{x}}\cdot\vec{\tilde{x}}=(t^2-\vec{x}\cdot\vec{x})(1-\frac{t_P}{t})^2
\end{eqnarray}
is invariant under the KMM transformation. The metric of the
coordinate system is then given by
\begin{eqnarray}
ds^2=d\tilde{t}^2-d\vec{\tilde{x}}\cdot
d\vec{\tilde{x}}=dt^2-[\frac{t_P}{t^2}\vec{x}dt+(1-\frac{t_P}{t})
d\vec{x}~]^2. \label{dskmm}
\end{eqnarray}
The explicit components of the metric tensor read
\begin{eqnarray}
g^{KMM}_{00} &=& 1-\frac{t_P^2~\vec{x}\cdot\vec{x}}{t^{4}} , \\
g^{KMM}_{0i} &=& -\frac{t_{P}x^{i}}{t^{2}}(1-\frac{t_{P}}{t}) , \\
g^{KMM}_{ij} &=& -\delta_{ij}(1-\frac{t_{P}}{t})^{2},
\end{eqnarray}
with its determinant given by
\begin{eqnarray}
g^{KMM}=-(1-\frac{t_{P}}{t})^{6}.
\end{eqnarray}~~
The maximum velocity of the signal propagation in DSR is
determined by the condition for the null geodesic,  $ds^2=0$, in
Eq. (\ref{dskmm}):
\begin{eqnarray}
[\vec{v}+(\frac{1}{1-\frac{t_P}{t}})\frac{t_P
\vec{x}}{t^2}]^2=\frac{1}{(1-\frac{t_P}{t})^2}.
\end{eqnarray}
Thus, in the direction where $\vec{x}$ is parallel to $\vec{v}$,
the maximum {\it signal velocity} is given by
\begin{eqnarray}
v=\frac{1}{1-\frac{t_P}{t}}(1-\frac{t_P x}{t^2}). \end{eqnarray}
This shows that the maximum signal velocity depends on the
position in space-time. The signal velocity at the origin,
$\vec{x}=0$, is infinitely large near the Planck time and
monotonically decreases in time to the conventional light velocity
when time approaches infinity. This is exactly what one expected
when DSR was first introduced for application to Planck-scale
physics [4,5,8].

To diagonalize the metric in Eq. (\ref{dskmm}), we define a new
radial coordinate $r$ as
\begin{eqnarray}
r=\sqrt{\vec{x}\cdot\vec{x}}~t(t-t_P), \end{eqnarray} which turns
the metric in Eq. (\ref{dskmm}) into
\begin{eqnarray}
ds^2=\frac{1}{t^4}[(\sqrt{t^4-\frac{4r^2}{t^2}}~dt+
\frac{2r}{t\sqrt{t^4-\frac{4r^2}{t^2}}}~dr)^2-
\frac{t^6}{t^6-4r^2}~dr^2-r^2(d\theta^2+\sin^2\theta
d\phi^2)].
\end{eqnarray}
If new variables $T$ and $\tilde{R}$ as
\begin{eqnarray}
T=t^2,~~\tilde{R}=\frac{2r^2}{T^3}
\end{eqnarray}
are introduced, the metric in Eq. (19) may be put into a diagonal
form:
\begin{eqnarray}
ds^2&=&\frac{1}{T^2}[\frac{Td\tau^2}{1-2\tilde{R}}-
\frac{dr^2}{1-2\tilde{R}}-r^2(d\theta^2+{\sin}^2\theta
d\phi^2)]\\
&=&\frac{1}{T(1-2\tilde{R})}[d\tau^2-\frac{1-2\tilde{R}}{T}dl^2],
\label{dskmm2}
\end{eqnarray}
where the new time variable $\tau$ is defined by
\begin{eqnarray}
\tau=\frac{1}{2}(T+\tilde{R}T),
\end{eqnarray}
and the metric of the space-like hypersection is given by
\begin{eqnarray}
dl^2=\frac{dr^2}{1-2\tilde{R}}+r^2(d\theta^2+\sin^2\theta
d\phi^2).
\end{eqnarray}
Although the diagonalized metric in Eq. (\ref{dskmm2}) is similar
in form to that of the Fock-Lorentz case {\cite {manida}}, it is
not a Friedmann-Lobachevsky type metric because the spatial part
of the metric, $dl^2$, is $\tau$-dependent. Note also that the
Planck-scale invariant $t_P$ does not appear in the metric in Eqs.
(19) and (21). Although the KMM transformation is singular at
$t=t_P$ and, thus, the physics before and after this time must be
treated separately, this implies that $t=t_P$ is not a
geometrically special point. Although this fact appears to be
unexpected, it is natural in view of the fact that the space-time
structure of doubly special relativity is obtained by a nonlinear
coordinate transformation, Eq. (9), from the Minkowski space-time;
thus, the curvature tensor obtained from the metric in Eq. (11)
vanishes.

We now consider the Fock-Lorentz(FL) transformation which is
induced by the transformation, $V=e^{-\frac{t}{R} D}$:
\begin{eqnarray}
t'&=&\frac{\gamma(t-\textit{v}x^{3})}{1-(\gamma-1)\frac{t}{R}+\gamma
\textit{v}\frac{x^{3}}{R}} , \\
x'^3&=&\frac{\gamma(x^{3}-\textit{v}t)}{1-(\gamma-1)\frac{t}{R}+\gamma
\textit{v}\frac{x^{3}}{R}} ,  \\
x'^a&=& \frac{x^{a}}{1-(\gamma-1)\frac{t}{R}+\gamma
\textit{v}\frac{x^{3}}{R}},  ~~~ a=1,2,
\end{eqnarray}
where R is a large invariant length scale, and the direction of
the relative motion between the two frames is assumed to be in the
$x^{3}$-direction. Note that, if one defines new variable
$\tilde{x}^{\mu}$ as
\begin{eqnarray}
\tilde{x}^{\mu}=\frac{x^{\mu}}{1+\frac{t}{R}}~, \label{xtilde1}
\end{eqnarray}
this new variable transforms as a 4-vector under the linear
Lorentz transformation, and the quantity
$\tilde{t}^{2}-\vec{\tilde{x}}\cdot\vec{\tilde{x}}
=(t^{2}-\vec{x}\cdot\vec{x})/{(1+\frac{t}{R})^{2}}$ can easily be
shown to be invariant under the Fock-Lorentz transfomation [1-2].
By using Eq. (\ref{xtilde1}), one finds the metric of the
coordinate system,
\begin{eqnarray}
ds^2=\frac{1}{(1+\frac{t}{R})^4}[dt^2-((1+\frac{t}{R})d\vec{x}-\frac{\vec{x}}{R}dt)^2].
\label{dsfl} \end{eqnarray} This metric may be put into a diagonal
form\cite{manida}: \begin{eqnarray}
ds^2=\frac{R^4}{\hat{t}^4}(d\hat{t}^2-\frac{\hat{t}^2}{R^2}dl^2),
\end{eqnarray}
where $t=\hat{t}\sqrt{1-\frac{r^2}{R^2}}$, and $dl^2$ is given by
\begin{eqnarray}
dl^2=\frac{1}{1-\frac{r^2}{R^2}}(\frac{dr^2}{1-\frac{r^2}{R^2}}+r^2(d\theta^2+\sin^2\theta
d\phi^2)).
\end{eqnarray}

The components of the metric tensor in Eq. (\ref{dsfl}) read
\begin{eqnarray}
g^{FL}_{00}&=&\frac{1-\frac{\vec{x}\cdot\vec{x}}{R^{2}}}{(1+\frac{t}{R})^{4}} ,\\
g^{FL}_{0i} &=&\frac{\frac{x^{i}}{R}}{(1+\frac{t}{R})^{3}} , \\
g^{FL}_{ij}&=&-\frac{\delta_{ij}}{(1+\frac{t}{R})^{2}},
\end{eqnarray}
whose determinant is given by
\begin{eqnarray}
g^{FL}=-(1+\frac{t}{R})^{-10}.
\end{eqnarray}
The maximum signal velocity in Fock-Lorentz nonlinear relativity
is determined by the null geodesic condition, $ds^2=0$, in Eq.
(\ref{dsfl}):
\begin{eqnarray}
[(1+\frac{t}{R})\vec{v}-\frac{\vec{x}}{R}]^2=1. \end{eqnarray}
Thus, in the direction where $\vec{x}$ is parallel to $\vec{v}$,
the maximum signal velocity becomes
\begin{eqnarray}
v=\frac{1}{1+\frac{t}{R}}(1+\frac{x}{R}).
\end{eqnarray}
As in the case of the KMM nonlinear relativity, the maximum signal
velocity in the FL nonlinear relativity depends on the position in
space-time. The velocity at the origin, $\vec{x}=0$, on the other
hand, is maximum and has the value of the conventional light
velocity at $t=0$ and monotonically decreases in time to become
zero when $t$ becomes infinite.

\section{Solutions of Doubly Special Relativistic Maxwell's Equations }

To understand the physical implications of doubly special
relativity, we solve Maxwell's equations invariant under the KMM
transformation\cite{kmm} and the Fock-Lorentz
transformation\cite{fock}. Maxwell's equations, invariant under
the nonlinear Lorentz transformations Eqs. (6)-(8) and Eqs.
(25)-(27), are of the form
\begin{eqnarray}
D_\alpha F^{\alpha\beta}&=& 4\pi J^\beta ,
\label{covf}\\
D_\alpha ~^\ast F^{\alpha\beta}&=&0 ,
\label{covfst}
\end{eqnarray}
where $D_\alpha$ represents a covariant derivative induced by the
metrics,  Eq. (11) and Eq. (29), for the KMM and FL  DSR,
respectively. Thus, Eqs. (\ref{covf}) and (\ref{covfst}) may be
written as
\begin{eqnarray}
\frac{1}{\sqrt{-g}}\frac{\partial(\sqrt{-g}F^{\alpha\beta})}{\partial
x^\alpha}&=&4\pi J^\beta ,
\label{covj}\\
\frac{1}{\sqrt{-g}}\frac{\partial(\sqrt{-g}~^\ast
F^{\alpha\beta})}{\partial x^\alpha}&=&0.
\label{cov0}
\end{eqnarray}

To solve Eqs. (\ref{covj}) and (\ref{cov0}), one finds it
convenient to write $F^{\alpha\beta}$  in terms of the electric
and magnetic fields, $E_i=F^{0i}, B_i=~^\ast F^{0i}$. Then, the
homogeneous equation, Eq. (\ref{cov0}), becomes
\begin{eqnarray}
\partial_i B_i&=&0, \\
D_0B_i+\epsilon_{ijk}\partial_j E_k&=&0,
\end{eqnarray}
where $D_0=\frac{1}{\sqrt{-g}}\partial_0\sqrt{-g}$, which reflects
the fact that the space-time coordinates are nonlinear. These
equations can be solved by writing the fields in terms of the
potential $A_\mu$;
\begin{eqnarray}
B_i&=&\epsilon_{ijk}\partial_j A_k ,
\label{bfield}\\
E_i&=&\partial_iA_0-D_0 A_i.
\label{efield}
\end{eqnarray}
The inhomogeneous equation, Eq. (\ref{covj}), becomes
\begin{eqnarray}
\partial_i E_i&=&4\pi J^0 ,
\label{gauss}\\
-D_0E_i+\epsilon_{ijk}\partial_jB_k&=&4\pi J_i.
\label{ampere}
\end{eqnarray}
Note that the system is invariant under the gauge transformation
\begin{eqnarray}
A_i\rightarrow A_i'&=&A_i+\partial_i \Lambda, \\
A_0\rightarrow A_0'&=&A_0+D_0 \Lambda. \end{eqnarray}

Substituting Eqs. (\ref{bfield}) and (\ref{efield}) into Eqs.
(\ref{gauss}) and (\ref{ampere}), one obtains the wave equation
for $A_\mu$:
\begin{eqnarray}
(D_0^2-\partial_i\partial_i)A_\mu=4\pi J_\mu,
\label{waveeq}
\end{eqnarray} where
we have used the gauge-fixing condition  $-D_0 A_0+\partial_i
A_i=0$. Note that Eq. (\ref{waveeq}) is equivalent to
\begin{eqnarray}
(\partial_0^2-\partial_i\partial_i)\tilde{A}_\mu=4\pi\tilde{J}_\mu,
\label{waveeqtilde}
\end{eqnarray} where
\begin{eqnarray}
\tilde{A}_\mu=\sqrt{-g}A_\mu, ~~\tilde{J}\mu=\sqrt{-g}J_\mu.
\label{aj}
\end{eqnarray}
One can now readily find the solutions of the doubly special
relativistic Maxwell's equations. In the source-free region, the
general solution is of the form
\begin{eqnarray}
A^\mu(t,\vec{x})=\frac{1}{\sqrt{-g}}\int d^3k
d\omega\epsilon^\mu(\omega,\vec{k})e^{i\vec{k}\cdot\vec{x}-i\omega
t},
\end{eqnarray}
where $\epsilon^\mu$ is the polarization vector satisfying
\begin{eqnarray}
\epsilon^0 \omega=\vec{\epsilon}\cdot\vec{k},
\end{eqnarray}
and the dispersion relation is given by
\begin{eqnarray}
\omega^2-k^2=0.
\end{eqnarray}
This shows that the solutions of Maxwell's equations are just
plane waves multiplied by an envelope function,
$\frac{1}{\sqrt{-g}}$, coming from the background metric factor
appearing in Eq. (\ref{aj}). Note that, for the KMM case, the
determinant of the metric is
\begin{eqnarray}
\sqrt{-g^{KMM}}=(1-\frac{t_P}{t})^3,
\end{eqnarray}
and, for the Fock-Lorentz case, it is
\begin{eqnarray}
\sqrt{-g^{FL}}=(1+\frac{t}{R})^{-5}.
\end{eqnarray}

Thus, in the case of the KMM DSR, the amplitude of the wave
decreases in time, starting from the Planck time when the
amplitude is infinitely large. In the Fock-Lorentz case, on the
other hand, the amplitude of the wave increases in time, starting
from $t=0$ when the amplitude is the same as it is in the
conventional plane-wave case. It appears that energy is
continuously extracted from (supplied to) the Maxwell system in
the KMM case (FL case), although there is no source or sink in the
system. Although, in the case of KMM, the Maxwell wave appears to
have infinitely large energy near the Planck time, the wave
velocity is not affected by the nonlinearity of the Lorentz group
representation. This is also the case for the Fock-Lorentz
relativity.

It is to be noted that the velocity of the Maxwell waves is the
same as that of the conventional light wave, but is not the same
as the maximum signal velocity, Eqs. (17) and (37) for the KMM and
the FL relativity, respectively, which follow from the null
geodesic condition. This is in marked contrast to the case of
linear relativity, where light propagates along the null
geodesics.

The fact that the energy appears to be supplied or extracted
without any source or sink in DSR can also be seen in Poynting's
theorem. From Maxwell's equations, Eqs. (\ref{gauss}) and
(\ref{ampere}), one obtains the continuity equation
\begin{eqnarray}
D_0u+\frac{1}{4\pi}\vec{\nabla}\cdot(\vec{E}\times\vec{B})=-\vec{J}\cdot\vec{E},
\label{poynting}
\end{eqnarray}
where
\begin{eqnarray}
u=\frac{1}{8\pi}\int d^3 x (E^2+B^2).
\end{eqnarray}
Note that the first term in Eq. (\ref{poynting}) can be written as
\begin{eqnarray}
D_0u=\partial_0u+\frac{1}{\sqrt{-g}}(\partial_0\sqrt{-g})~u.
\label{d0}
\end{eqnarray}
The second term of Eq. (\ref{d0}) may act like a source term, and
due to this term, the energy-momentum of the wave is not conserved
globally.

\section{Discussion}

DSR in momentum space is obtained by defining the modified algebra
[4,5,8,11]
\begin{eqnarray}
K^i=U^{-1} L_0^{~\:i}~U,
\end{eqnarray}
where ~$L_{\alpha\beta}=p_\alpha\frac{\partial}{\partial
p^\beta}-p_\beta\frac{\partial}{\partial p^\alpha}$~ are the
generators of the linear Lorentz transformation. From the modified
generators, $K^i$, one obtains the modified Lorentz transformation
in momentum space, which in turn gives the modified dispersion
relation
\begin{eqnarray}
E^2 f^2(E)-p^2g^2(E)=m^2,
\label{disp}
\end{eqnarray}
where the forms of the functions \textit{f} and \textit{g} are
determined by the form of the transformation $U$. From the
dispersion relation in Eq. (\ref{disp}), one defines the group
velocity of the particle in question and discusses the effects in
Planck-scale physics [7].

Although  \textit{p} in the linear Lorentz generators,
$L_{\alpha\beta}$, is clearly the momentum of the particle, it is
not clear what represents the variable  \textit{p} appearing in
the modified generators, $K^i$, and in the nonlinear Lorentz
transformation. Thus, it is not clear whether the ``group
velocity" defined from the dispersion relation in Eq. (\ref{disp})
represents the true group velocity of the particle in question. We
believe that this fact is the source of the questions raised by
the authors of Refs. [9,10].

One way to define the velocity of a particle unambiguously is to
solve the field equation corresponding to the particle in
question. Since the exact nonlinear Lorentz transformation in
position space corresponding to the momentum-space DSR relevant in
Planck-scale physics [4,5]
is not yet known, we solved Maxwell's equations invariant under
the KMM transformation\cite{kmm} and the Fock-Lorentz
transformation\cite{fock} in an attempt to understand how the
light wave propagates in DSR. As we showed in the last section,
although the Maxwell wave in KMM nonlinear relativity appears to
acquire infinitely large energy near the Planck time, its velocity
is the same as that of conventional light waves. The reason is
that the nonlinear Lorentz transformation does not affect the
phase of the Maxwell wave, but contributes only to the amplitude
of the wave.

Another puzzle present in doubly special relativity defined in
position space is that the velocity of the doubly special
relativistic Maxwell wave is not the same as the maximum signal
velocity determined by the null geodesics. The maximum signal
velocity in KMM nonlinear relativity is infinitely large near the
Planck time and approaches the conventional light velocity as time
becomes large, which is exactly the property that is needed for
application to Planck-scale physics. The velocity of the solution
of Maxwell's equations invariant under the KMM transformation,
however, is exactly the same as the conventional light velocity,
although its energy becomes infinitely large near the Planck time.
This raises many interesting questions, such as the one concerning
the relation between null geodesics and the particle trajectory or
wave propagation, whether there exists a particle or a wave that
travels with the maximum signal velocity determined by the null
geodesics, {\it etc}. This also raises another serious question on the
causality of the theories based on the KMM transformations because
the light cone determined by Maxwell's equations may be inside or
outside the null cone determined by the null geodesic equation,
Eq. (16), depending on the position of the particle in question.
Note that the dispersion relation, Eq. (62), is obtained from the
metric of the coordinate system defined in momentum space.
However, our analysis indicates that the signal velocity
determined by the metric may not be the same as that determined by
the solutions of the field equations. Thus, if the physics of
doubly special relativity are to be understood more clearly,
further investigations of the field equations invariant under the
nonlinear Lorentz transformations and the implications of the null
geodesics are needed.

\begin{acknowledgments}
This work was supported in part by the Sabbatical Year Program of
Ewha Woman's University (SKK), by the Yonsei University Research
Fund (SMK), by the Basic Research Program of the Korea Science and
Engineering Foundation under Grant number
R01-1999-000-00018-0(2003)(CR), and by Korea Research Foundation
under Project number KRF-2003-005-C00010 (JHY).
\end{acknowledgments}

\end{document}